\documentstyle[12pt,epsfig]{article}
\def\agt{\stackrel{\displaystyle >}{\sim}}

\begin{document}

\noindent TPR-97-03, ULM--TP/97-5 \\
April 1997
\vspace{3.0cm}

\centerline{\LARGE Semiclassical Interpretation}
\vspace{0.5cm}
\centerline{\LARGE of the Mass Asymmetry in Nuclear Fission}
\vspace{2.0cm}
\centerline{\large M.\ Brack$^1$, S.\ M.\ Reimann$^2$ and 
M.\ Sieber$^3$}
\vspace{0.4cm}

\noindent $^1$ Institut f\"ur Theoretische Physik, Universit\"at
               Regensburg, D-93040 Regensburg, Germany \\ 
\noindent $^2$ Niels Bohr Institutet, Blegdamsvej 17, DK-2100
               Copenhagen \O, Denmark \\
\noindent $^3$ Abteilung Theoretische Physik, Universit\"at Ulm,
               D-89069 Ulm, Germany

\vspace{2.4cm}
\centerline{\bf Abstract}
\vspace{0.5cm}

We give a semiclassical interpretation of the mass asymmetry
in the fission of heavy nuclei. Using only a few classical periodic orbits and
a cavity model for the nuclear mean field, we reproduce the onset of
left-right asymmetric shapes at the fission isomer minimum and the correct
topology of the deformation energy surface in the region of the outer fission
barrier. We point at the correspondence of the single-particle quantum states
responsible for the asymmetry with the leading classical orbits, both
lying in similar equatorial planes perpendicular to the symmetry axis of the
system.

\vspace{2.5cm}

\noindent PACS numbers: \\
\noindent 24.75.+i ~ General properties of fission \\
\noindent 03.65.Sq ~ Semiclassical theories and applications. \\
\noindent 47.20.Ky ~ Nonlinearity (including bifurcation theory).

\vspace{0.3cm}

\newpage

One characteristic feature of the fission of many actinide nuclei is the
asymmetric mass distribution of the fission fragments. The liquid drop model
\cite{ldmbw}, although able to describe many aspects of the fission process
qualitatively, cannot explain this mass asymmetry in heavy nuclei where the
fissility parameter $x$ is close to unity \cite{wilet}: the balance between the
attractive surface tension and the repulsive Coulomb force favors left-right
symmetric shapes and thus also the symmetric fission. An explanation for the
observed asymmetry of the fragment distributions became possible with
Strutinsky's shell-correction method \cite{strut} which includes
the quantal shell effects stemming from the discrete spectra of
the nucleons in their mean fields. One writes the total binding
energy of a nucleus consisting of $N$ neutrons and $Z$ protons as
\begin{eqnarray}
E_{tot}(N,Z;def) & = & E_{LDM}(N,Z;def) \nonumber \\
                 &   & + \delta E_n(N;def) + \delta E_p(Z;def)\,. \label{Enuc}
\end{eqnarray}
Here $E_{LDM}$ is the liquid drop model (LDM) energy; $\delta
E_n$ and $\delta E_p$ are the shell-correction energies of the neutrons and
protons, respectively, which are obtained in terms of the single-particle
energies of realistic shell-model potentials. All ingredients depend on the
shape of the nucleus, which is described by some suitable deformation
parameters, summarized in (\ref{Enuc}) by `$def$'. The shell-correction
approach was very successful in reproducing experimental nuclear binding
energies and fission barriers \cite{fuhil,nilss,bolfi} at times where purely
microscopical selfconsistent calculations of Hartree-Fock type were not yet
available \cite{hafo}.

\begin{figure}
\begin{center}
\mbox{\epsfig{file=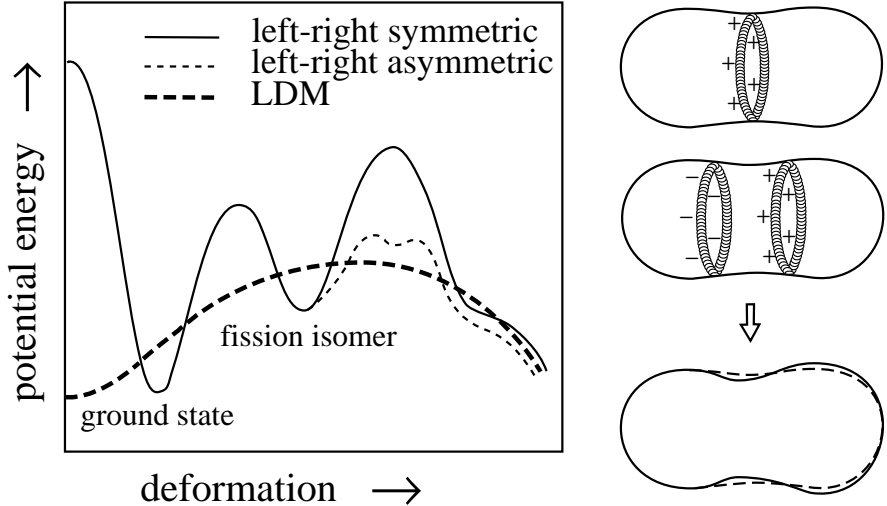,width=12cm}}
\end{center}
\caption{
{\it Left side}: schematic fission barrier of a typical actinide nucleus (after
Ref.\ \protect\cite{mbjue}). {\it Right side}: schematic probability amplitudes
of the leading single-particle wavefunctions responsible for the onset of mass
asymmetry (after Ref.\ \protect\cite{gumni}).
}
\label{sysbar}
\end{figure}

In Fig.\ \ref{sysbar} we show a schematic fission barrier of a typical actinide
nucleus, taken along the adiabatic path through the multi-dimensional
deformation space. The heavy dashed line is the LDM deformation energy which
leads to a spherical ground state and to equal fragment sizes. The solid line
is the total energy according to Eq.\ (\ref{Enuc}); it contains the typical
shell oscillations coming from the shell corrections $\delta E_n$ and $\delta
E_p$. These lead to a deformed ground state minimum and to a higher-lying local
minimum, the so-called fission isomer. (For an extended review on the physics
of the `double-humped fission barrier', see Ref.\ \cite{bjoly}.) The shapes
assumed hereby have axial symmetry and left-right symmetry. When the latter
is relaxed, the energy is found \cite{moeni,paleb} to be lowered along
the way over the outer barrier, starting at the fission isomer. The gain in
energy persists all the way down towards the scission point, where the nucleus
breaks into two fragments of unequal size. Nonaxial deformations do not change
this feature; they only lead to a slight reduction of the inner barrier (see
Ref.\ \cite{mbjue} for a short review of fission barrier calculations and the
role of various deformations). It is important to note that the onset of the
mass asymmetry takes place already at an early stage of the fission
process, long before the nucleus breaks up. It is a pure quantum effect which
only comes about if the shell corrections are included into the total energy.
The microscopic origin of this instability has been investigated by Gustafsson,
M\"oller, and Nilsson \cite{gumni}. They found that only two specific types of
single-particle states with large angular momenta along the symmetry axis are
strongly sensitive to the left-right asymmetry: one of them has the maxima of
its wave functions along the central waist line of the nucleus (see
the upper right in Fig.\ \ref{sysbar}), whereas the other has maxima along
the circumferences of two equatorial planes at some distance of the center
(with opposite phases on either side, see the middle right in Fig.\
\ref{sysbar}). The coupling of these states through the left-right asymmetric
components of the mean field leads to a decrease of one set of eigenenergies
which lie below the Fermi energy, and thus to a reduction of the total binding
energy when the asymmetry is switched on (see the lower right in Fig.\
\ref{sysbar}).

In the following we give a very simple and transparent semiclassical
interpretation of this quantum effect. We employ the periodic orbit theory
(POT) developed independently by Gutzwiller \cite{gutzw} for systems with
isolated classical orbits and by Balian and Bloch \cite{bablo} for arbitrarily
deformed cavities with reflecting walls. In the POT, one obtains the
oscillating part of the level density of a quantum system from the so-called
`trace formula':
\begin{equation}
\delta g\,(E) \simeq \sum_{po} {\cal A}_{po} (E) \cos \left[ \frac{1}{\hbar}
                   S_{po} (E) - \sigma_{po} \frac{\pi}{2} \right].  \label{tf}
\end{equation}
The sum is taken over all periodic orbits, labeled `$po$', of the corresponding
classical system. $S_{po}$ are the classical actions along the periodic orbits
and $\sigma_{po}$ are phases related to the number of conjugate points along the
orbits. The amplitude ${\cal A}_{po}$ of each orbit depends on its period, its
stability and its degeneracy. Together with the smooth part ${\widetilde
g}\,(E)$ which can be obtained in the (extended) Thomas-Fermi model, it
approximates the exact quantum-mechanical level density: ${\widetilde g}\,(E) +
\delta g\,(E) = g\,(E) = \sum_i \delta(E-E_i)$, where $E_i$ are the
eigenenergies of the system and the sum runs over all quantum states $i$. (See
Ref.~\cite{brbha} for an introduction to the POT and detailed explanations of
all the above ingredients). Gutzwiller's trace formula has initiated promising
steps towards the semiclassical quantization of chaotic systems \cite{pot}.
A different use of the POT \cite{brbha,strma} is to obtain a coarse-grained
level density by keeping only the shortest orbits with the largest amplitudes
in the trace formula (\ref{tf}). This allows one to relate the gross-shell
structure of interacting fermion systems in the mean-field approximation to a
few classical orbits. Using Eq.\ (\ref{tf}), the semiclassical expression
for the energy shell correction $\delta E$ becomes \cite{strma}
\begin{equation}
\delta E \simeq \sum_{po} {\cal A}_{po} (E_F) \!
                \left(\frac{\hbar}{T_{po}}\right)^{\!2} \!
                \cos \left[ \frac{1}{\hbar}
                S_{po} (E_F) - \sigma_{po} \frac{\pi}{2} \right].  \label{Esc}
\end{equation}
Here $E_F$ is the Fermi energy and $T_{po}=dS_{po}/dE|_{E_F}$ the
period of the
orbit labeled $po$. Pioneering work has been done in this direction by
Strutinsky {\it et al.}~\cite{strma}, who generalized the Gutzwiller theory to
systems with continuous symmetries and used it to give a semiclassical
explanation of the systematics of nuclear ground-state deformations. They
showed that the ground-state minima of the nuclear binding energy as functions
of particle number and deformation follow the loci of constant action of the
shortest periodic orbits. Another beautiful example is the beating pattern of
the coarse-grained level density in a spherical cavity, which was related by
Balian and Bloch \cite{bablo} to the interference of the triangular and square
periodic orbits, and later predicted \cite{nishi} and observed \cite{supsh} in
metal clusters in the form of the so-called supershells.

For our semiclassical investigation, we replace the nuclear mean field by a
cavity with reflecting walls and consider only one kind of particles. This
should yield the main physical effects, since the neutron contribution $\delta
E_n$ to the total energy contains the largest part of the shell oscillations,
as seen, e.g., in Fig.\ VIII-4 of Ref.\ \cite{fuhil}. We employ the
parameterization ($c,h,\alpha$) of this reference, used there to define the
shape of the liquid drop. Here it defines the boundary of the cavity in
cylindrical coordinates $(\rho,z,\phi)$, with the symmetry axis along $z$, by
the shape function $\rho=\rho(z;c,h,\alpha)$. $c$ is the length of the semiaxis
along the $z$ direction in units of the radius $R_0$ of the spherical cavity
($c$=1, $h$=$\alpha$=0). $h$ is a `neck parameter' that regulates the formation
of a neck leading to the scission of the nucleus into two fragments.
$\alpha\neq 0$ yields left-right asymmetric shapes. The volume of the cavity is
kept constant for all deformations. (See Ref.\ \cite{fuhil} for the details of
this parameterization, and especially Figs.\ VII-1 and VIII-5 for the most
important shapes occurring in fission.) The parameters $(c,h)$ are chosen such
that the one-dimensional curve $h$=$\alpha$=0 along $c$ follows the adiabatic
fission barrier of the LDM (shown schematically in Fig.\ \ref{sysbar}). Even
including the shell effects, $h$=0 gives a reasonable picture of the
double-humped fission barrier.

We now have to determine the shortest periodic orbits of this system to
calculate the gross-shell structure in $\delta E$. At large deformations (here
$c\agt 1.4$), these are the orbits lying in equatorial planes perpendicular to
the symmetry axis \cite{frisk,monst}. The positions $z_i$ of these planes
along the $z$ axis are given by the condition that the shape function be
stationary: $d\rho(z;c,h,\alpha)/dz|_{z_i}=0$. The periodic orbits have the
form of regular polygons and are characterized by $(p,t)$, where $p$ is the
number of reflections at the boundary and $t$ the number of windings around the
symmetry axis. Both numbers are restricted by $p \geq 2$ and $t \leq p/2$. The
semiclassical contribution of such orbits to the trace formula (\ref{tf}) has
been derived by Balian and Bloch \cite{bablo}; we refer to their paper for the
explicit form of the amplitudes ${\cal A}_{pt}$ and phases $\sigma_{pt}$. The
lengths of the orbits are $L_{pt}^{(i)} = 2 p R_i \sin(\pi t/p)$, where
$R_i=\rho(z_i;c,h,\alpha)$, so that their actions equal
$S_{pt}^{(i)}(E_F)=\hbar k_F L_{pt}^{(i)}$ in terms of the Fermi wave
number $k_F=\sqrt{2mE_F}/\hbar$.

The range of validity of Eqs.\ (\ref{tf}) and (\ref{Esc}) is, however, limited.
They are correct only as long as the orbits are well separated from neighboring
periodic orbits, in particular as long as the orbits are not close to a
bifurcation. At a bifurcation the amplitudes $A_{pt}$ diverge and the trace
formula has to be modified. Generally, bifurcations exist in 
different forms, but for the
shapes studied here we need consider only one type of bifurcation.
It occurs when the positions $z_i$ of several equatorial planes coincide. In
the $(c,h,\alpha)$ parameterization, there are at most three such planes. One
plane always exists; the other two arise at
the points $(c_0,h_0,\alpha_0)$ where the neck formation starts. In
the symmetric case ($\alpha$=0), one plane is always located at $z_0$=0 and,
beyond the bifurcation point, the other two are located symmetrically at $\pm
z_1$ (with $z_1$$>$0) and contain identical periodic orbits.

Near a bifurcation point, the neighboring orbits give a joint contribution to
the level density. This contribution can be derived from an oscillatory
integral which contains the contributions of all orbits of type ($p,t$) from
the different planes. The integral is of the form
\cite{bablo}
\begin{equation}
\delta E_{pt} = \Re e\, \int_{-c}^{+c} \! dz \,
f_{pt}(z) \exp \{ i k_F L_{pt}(z) \} \, ,                      \label{intosc}
\end{equation}
where $f_{pt}(z)$ are slowly varying analytic functions of $z$. Since the plane
positions $z_i$ of the periodic orbits are determined by the stationary points
of the length function $L_{pt}(z) = 2 p \sin(\pi t/p) 
\cdot \rho(z;c,h,\alpha)$, a
stationary phase evaluation of (\ref{intosc}) leads back to separate 
contributions to Eq.\ (\ref{Esc}) for each plane, with the amplitudes and
phases given in \cite{bablo}. In order to obtain an approximation to
(\ref{intosc}) that is valid at the bifurcation as well as far from it, we
employ a uniform approximation that is appropriate for the case of three nearly
coincident stationary points in a one-dimensional oscillatory integral
\cite{Con73}. It is expressed in terms of Pearcey's integral and its
derivatives:
\begin{eqnarray}
\delta E_{pt} & = & \Re e\, \left\{ \left[
                   u_4 P(u_1,u_2) + u_5 P_x(u_1,u_2) \right.\right.\nonumber \\
         &   & \qquad \left.\left. \qquad  + u_6 P_y(u_1,u_2)
                       \right] e^{i u_3} \right\} \, ,        \label{uniform}
\end{eqnarray}
where Pearcey's integral is defined by
\begin{equation}
P(x,y) = \int_{-\infty}^\infty \! dz \,
         \exp [\,i(z^4+x z^2 + y z)],                          \label{pearcey}
\end{equation}
and $P_x(x,y)$, $P_y(x,y)$ denotes its derivatives with respect to its first
and second argument, respectively. The real constants $u_1 \dots u_6$ are
determined by the semiclassical amplitudes, actions and phases of the periodic
orbits. If the orbits are well separated, Eq.\ (\ref{uniform}) reduces
to contributions
to the standard trace formula (\ref{Esc}).

When the asymmetry parameter $\alpha$ is sufficiently large, the contributions
from one plane of orbits can be evaluated in the stationary phase
approximation. Both in this limit and in the symmetric case ($\alpha$=0), the
result can be expressed in terms of cylindrical Bessel functions $J_\mu(x)$,
and the formulae are analogous to those for generic saddle-node and pitchfork
bifurcations \cite{SS97}. We give here the result for $\delta E_{pt}$ in
the symmetric case: 
\begin{eqnarray}
\delta E_{pt} & = &  \Re e \, \left[ 
                \sqrt{\frac{\pi k_F |\Delta L_{pt}|}{2}}
                e^{\,i \left(\!k_F\bar{L}_{pt}-3p\pi/2 \right)}
                \right. \nonumber \\ && \times
                \left\{ \left( \frac{\hbar^2{\cal
                A}_{pt}^{(1)}}{\left[T_{pt}^{(1)}\right]^2}
                + \frac{\hbar^2{\cal
                A}_{pt}^{(0)}}{\sqrt{2}\left[T_{pt}^{(0)}\right]^2} \right)
                \left(\nu \,J_{ 1/4}(k_F |\Delta L_{pt}|) \, e^{ i\pi/8}
                     + J_{-1/4}(k_F |\Delta L_{pt}|) \, e^{-i\pi/8} \right)
               \right.      \\ \nonumber
              &    & \left. \left.
              + \left( \frac{\hbar^2{\cal
                A}_{pt}^{(1)}}{\left[T_{pt}^{(1)}\right]^2}
                - \frac{\hbar^2{\cal
                A}_{pt}^{(0)}}{\sqrt{2}\left[T_{pt}^{(0)}\right]^2}
                \right)
              \left( J_{ 3/4}(k_F |\Delta L_{pt}|) \, e^{ i 3\pi/8}
                   + \nu \, J_{-3/4}(k_F |\Delta L_{pt}|) \, e^{-i 3\pi/8}
                \right) \right\} \right].
                                                                \label{unisym}
\end{eqnarray}
Here $\bar{L}_{pt} =
[L_{pt}^{(1)} + L_{pt}^{(0)}]/2$ and $\Delta L_{pt} = [L_{pt}^{(1)} -
L_{pt}^{(0)}]/2$ in terms of the lengths $L_{pt}^{(0)}$, $L_{pt}^{(1)}$ of the
orbits $pt$ situated at $z=z_0$ and $z=\pm z_1$, respectively. $\nu$ equals
$-1$ before the bifurcation (i.e., for only one orbit plane) and $+1$
after the bifurcation (for three orbit planes).

\begin{figure}
\begin{center}
\mbox{\epsfig{file=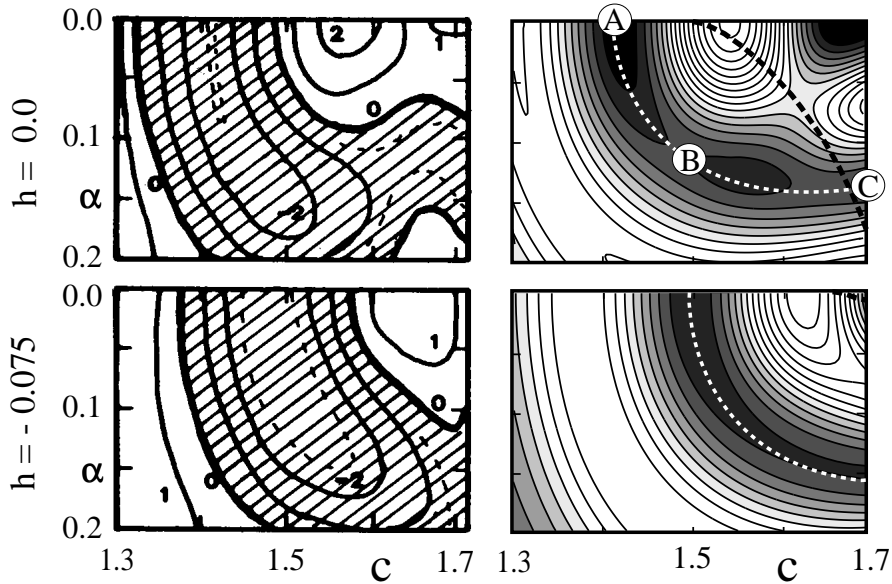,width=12cm}}
\end{center}
\caption{
Contour plots of deformation energy versus elongation $c$ and asymmetry
$\alpha$ for two values of the neck parameter: $h$=0 (above) and $h$=$-0.075$
(below). {\it Left side}: quantum-mechanical neutron shell-correction energy
of $^{240}$Pu from Ref.\ \protect\cite{fuhil}. {\it Right side}:
semiclassical shell-correction energy $\delta E$. White dashed lines indicate
the loci of constant classical action of the central equatorial
periodic orbits; black dashed lines the loci of the bifurcation points.
}
\label{contours}
\end{figure}

In the right-hand panels of Fig.\ \ref{contours} we show contour plots of the
semiclassical shell-correction energy $\delta E$ in the ($c,\alpha$) plane for
two values of the neck parameter. The energy unit is $E_0=\hbar^2\!/2mR_0^2$,
where $R_0$ is the radius of the spherical box. The Fermi wave number $k_F$ was
chosen such that $\delta E$ has a minimum at the deformation $c=$1.42,
$h$=$\alpha$=0 of the fission isomer. On the left of Fig.\ \ref{contours} we
have reproduced the neutron shell-correction energy $\delta E_n$
of the nucleus $^{240}$Pu, obtained in Ref.\ \cite{fuhil} with a realistic
Woods-Saxon type shell-model potential. We see that the semiclassical result
correctly reproduces the topology of the deformation energy in the $(c,\alpha)$
plane for both values of $h$, in particular the onset of the mass asymmetry at
the fission isomer. It should be noted that we have only included orbits with
winding number one ($t$=1) and with up to $p_{max}=10$ reflections. The results
for $\delta E$ are the same within a few percent when only orbits with $p=2$
and 3, i.e., only the diameter and triangle orbits, are included. The loci of
the bifurcation points $(c_0,\alpha_0)$ are indicated in Fig.\ \ref{contours}
by the black heavy dashed lines (that for $h$=$-0.075$ is hardly visible in the
upper right corner of the plot). This shows that the essential feature, namely
the energy gain due to the asymmetric deformations for $c\geq 1.42$, is brought
about by only two classical orbits: the diameter and the triangle in the
central equatorial plane. The white dashed lines give the loci of constant
actions of the periodic orbits at $z_0$, fixing their value at $\alpha$=0.
(Note that the actions of all orbits in a given equatorial plane have the same
deformation dependence.) We see that the valley that leads from the isomer
minimum over the outer fission barrier in the energetically most favorable way
is following exactly the path of constant action of the leading classical
orbits; the path is practically identical with that obtained in the
quantum-mechanical shell-correction calculations. A more detailed
comparison with the latter, together with the calculational details of
our studies, will be presented in a forthcoming publication.

\begin{figure}
\begin{center}
\mbox{\epsfig{file=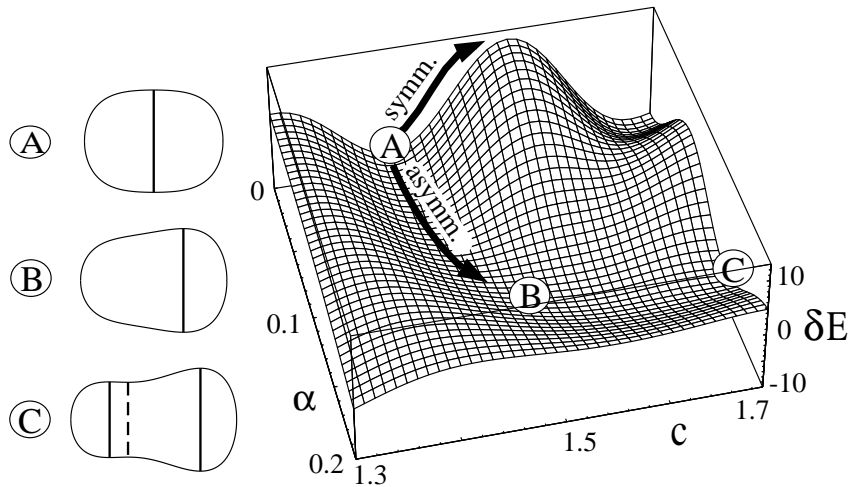,width=11.5cm}}
\end{center}
\caption{
{\it Right side}: semiclassical shell-correction energy $\delta E$ as in Fig.\
\protect\ref{contours} for $h=0$, but in a three-dimensional perspective view.
Arrows `symmetric' and `asymmetric' show two alternative fission paths. {\it
Left side}: shapes found along the asymmetric fission path; the
equatorial planes of periodic orbits included in the trace formula are
shown by the vertical lines (solid lines for stable and dashed lines for
unstable orbits).
}
\label{barview}
\end{figure}

In Fig.\ \ref{barview}, we show the same results as in the upper right part of
Fig.\ \ref{contours}, but in a perspective view of a three-dimensional energy
surface. On the left, the shapes of the cavity are given corresponding to the
points A at the isomer minimum and the points B and C along the asymmetric
fission barrier (see also the corresponding points in Fig.\ \ref{contours}).
Note that C lies beyond the bifurcation point and thus contains three planes of
periodic orbits.

In summary, we have shown how a specific quantum effect, causing a drastic
rearrangement of the shape of a complex many body system, can be described
semiclassically by the constancy of the actions of a few periodic orbits. We
point out the close correspondence of the equatorial planes of these orbits
(see the left of Fig.\ \ref{barview}) with the locations of the wave function
maxima (see the right of Fig.\ \ref{sysbar}) of the relevant quantum
states. 
We also note that the classical dynamics of this system is rather chaotic,
particularly in the asymmetric case and for small values of the conserved
angular momentum $L_z$ \cite{arima}. The regular regions in phase
space, connected to the stable periodic orbits with axial degeneracy,
are still strong enough to cause this important shell effect.

M.\ S.\ wishes to acknowledge financial support by the Deutsche
Forschungsgemeinschaft under Contract No.\ DFG-Ste 241/6-1 and /7-2.

\end{document}